\documentclass[conference]{IEEEtran}
\IEEEoverridecommandlockouts
\usepackage{xcolor,soul,framed} 
\usepackage{graphicx}
\usepackage[cmex10]{amsmath}
\usepackage{array}
\usepackage{mdwmath}

\usepackage{eqparbox}
\usepackage{url}
\usepackage{xurl}
\usepackage{amssymb}
\usepackage{orcidlink}
\usepackage{breqn}
\usepackage{float}
\usepackage{dblfloatfix}
\usepackage{subcaption}
 \usepackage{adjustbox}
\usepackage{booktabs}
\usepackage{multirow} 
\usepackage{enumitem}
\usepackage{tabularx}
\usepackage{makecell}
\usepackage{blindtext}
\usepackage{hyperref}
\usepackage{diagbox}
\usepackage{amsmath} 
\usepackage{mathtools}
\usepackage{lipsum}
\usepackage{pdfrender}
\usepackage[
  separate-uncertainty = true,
  multi-part-units = repeat
]{siunitx}
\usepackage{textcomp}
\usepackage{caption}
\usepackage{rotating}

\usepackage[backend=biber, style=ieee, citestyle=ieee-comp, maxbibnames=5, mincitenames=1]{biblatex}
\AtEveryBibitem{\setlength{\itemsep}{2pt}} 
\AtBeginBibliography{\footnotesize}
\usepackage{soul,color}

\usepackage{multirow}

\usepackage{supertabular}
\renewcommand{\arraystretch}{0.9}
\usepackage[T1]{fontenc}

\usepackage{threeparttable}  

\usepackage{pifont}

\usepackage{bbding}
\usepackage{wasysym}

\usepackage{cleveref}

\usepackage{makecell} 

\newif\ifshowmods
\showmodsfalse   

\ifshowmods

\else

\fi

 \makeatletter
\newcommand*\bigcdot{\mathpalette\bigcdot@{.5}}
\newcommand*\bigcdot@[2]{\mathbin{\vcenter{\hbox{\scalebox{#2}{$\m@th#1\bullet$}}}}}
\makeatother

\begin{document}

\title{
Comparison of Open-Source PSCAD EMT Models for Large-Scale AI Data Centers in Smart Grids
}

\author{
\IEEEauthorblockN{Alex Chen\orcidlink{0009-0003-0776-3887},
Ahmad Mohammad Saber\orcidlink{0000-0003-3115-2384}, and
Deepa Kundur\orcidlink{0000-0001-5999-1847} 
\IEEEauthorblockA{Department of Electrical and Computer Engineering, University of Toronto, Toronto, ON, Canada}
}
}

\maketitle

\begin{abstract}

The rapid deployment of artificial intelligence data centers (AIDCs) introduces gigawatt-scale, highly nonlinear, and fast-varying electronic loads that challenge grid voltage regulation, frequency stability, and disturbance ride-through. Capturing their transient behavior during grid faults and rapid computational workload swings requires high-fidelity electromagnetic transient (EMT) simulations rather than root mean square models. However, limited methodologies exist to audit the component-level adequacy of EMT AIDC models. This paper proposes a software-agnostic, 41-point fidelity framework spanning 11 categories that synthesizes 37 criteria from an existing modeling guideline with four novel requirements: frequency ride-through compliance, disaggregated fan and chiller cooling loads, temperature-dependent thermal response, and DC information technology (IT) load representations. Applying this framework to the four publicly available AIDC models in PSCAD/EMTDC reveals varying degrees of completeness: while the models capture IT load transitions, uninterruptible power supplies, and transfer switches for bulk stability studies, critical gaps remain in electro-thermal cooling coupling, internal protective relaying, and site-level controls. Addressing these gaps will enable comprehensive protection coordination and ride-through compliance studies.

\end{abstract}

\section{Introduction}
\label{sec: intro}

The rapid growth of artificial intelligence (AI) infrastructure and large computational loads (LCLs) is reshaping modern power system dynamics. Unlike traditional industrial facilities modeled via static or induction motor (IM) representations, AI data centers (AIDCs) present highly nonlinear, fast-varying electronic loads sensitive to grid disturbances \cite{talukdarModelingFrameworkData2026}. The Electric Reliability Council of Texas (ERCOT) defines LCLs as facilities with at least 75 MW of peak demand, 50\% or more of which comprises power-electronic computational equipment \cite{ercotNOGRR282LargeComputational2026}, a classification encompassing most AIDCs. Accurately capturing AIDC–grid interactions is therefore critical for reliable network planning and operation.

With single-site computational demand exceeding 1,037 MW at facilities such as New Carlisle \cite{epochaiAIDataCenters2026}, abrupt AIDC disconnections can trigger severe voltage and frequency excursions. During a net 3800 MW load transfer event on Virginia's grid, an initial 2970 MW data center disconnection induced voltage and frequency spikes that precipitated a further 1099 MW drop (Fig. 1) \cite{wharton2026dominion}. Consequently, grid operators are now mandating or proposing strict voltage and frequency ride-through (VRT/FRT) standards for LCLs \cite{ercotNOGRR282LargeComputational2026}, \cite{wharton2026dominion}.

Evaluating ride-through compliance and converter-driven transients requires electromagnetic transient (EMT) simulations, as phasor-domain root mean square (RMS) models neglect sub-cycle switching dynamics and high-frequency non-linearities, \cite{shahriarpramanikAssessingDynamicBehaviors2025}, \cite{favuzzaComparativeModelingAnalysis2024}. Transmission operators now require EMT studies for LCL interconnection requests \cite{ieso2026largecomputational}. Despite this need, only four open-source PSCAD EMT AIDC models are publicly available as of writing \cite{ERCOTModel}, \cite{NLRModel}, \cite{PNNLModel}, \cite{DartmouthModel}. To support model benchmarking and development, this paper introduces a 41-point, software-agnostic EMT AIDC fidelity framework and applies it to these four open-source models to identify their strengths and priority areas for extension.

\section{AI Data Center Topology}
 \label{section:Data Center Design}

Fig. \ref{fig:AIDC config} illustrates an AIDC comprising four primary subsystems: information technology (IT) loads (CPUs, GPUs, and networking hardware), cooling systems, an uninterruptible power supply (UPS) with battery energy storage (BESS), and auxiliary facility loads (lighting and site management). Below, we outline each subsystem's characteristics and explain why EMT modeling is required to capture its grid interactions.

\begin{figure}[t!]
    \centering
    \includegraphics[width=\columnwidth]{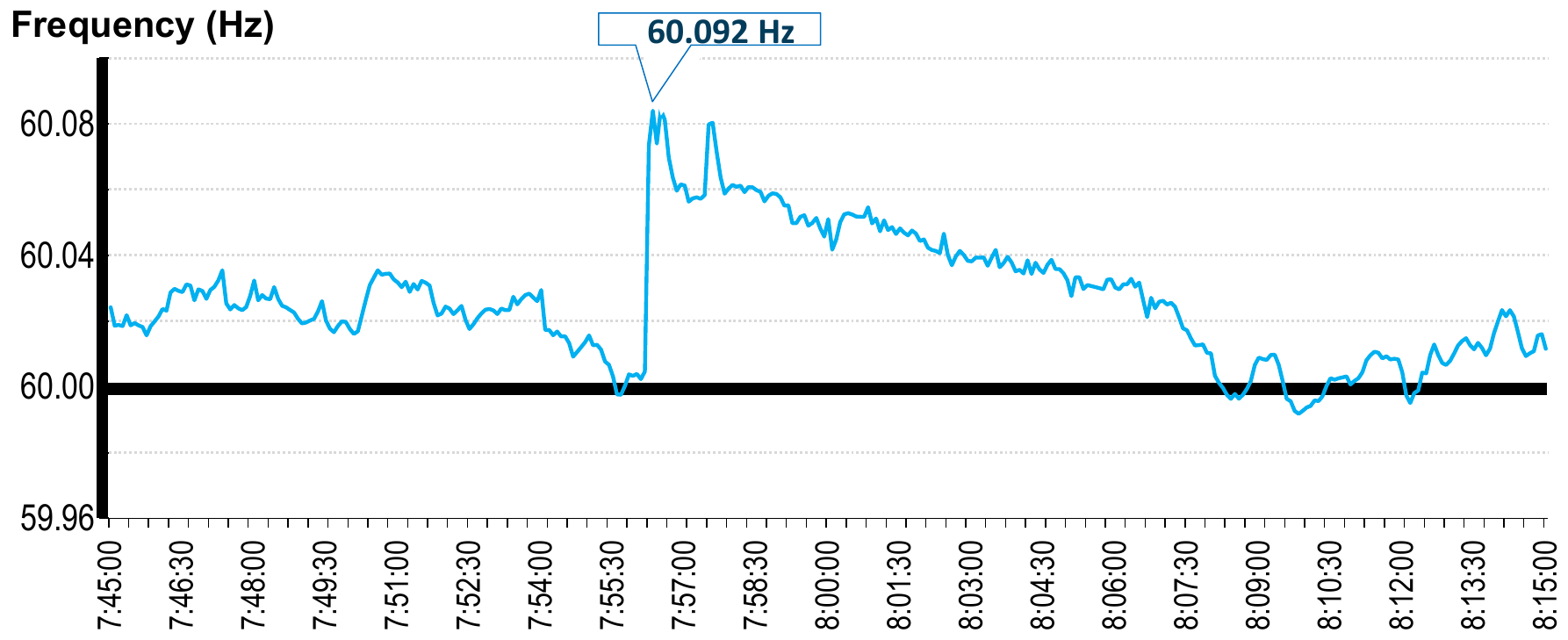}
    \caption{Grid frequency during a load transfer event \cite{wharton2026dominion}.}
    \label{fig: load transfer}
\end{figure}
\begin{figure*}[t!]
    \centering
    \includegraphics[width=\linewidth]{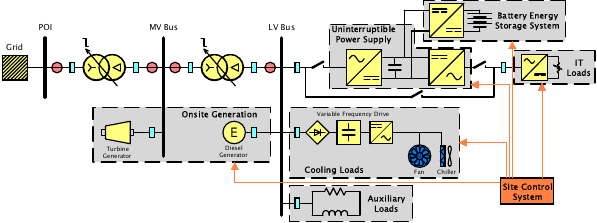}
    \caption{AIDC design with a centralized UPS.} 
    \label{fig:AIDC config}
\end{figure*}
\subsection{Information Technology Loads}

IT equipment is the dominant power consumer in an AIDC and exhibits highly volatile demand across two primary workloads: AI model training and inference \cite{koWideareaPowerSystem2026}. Training maintains high average utilization punctuated by abrupt, facility-wide power swings during periodic checkpointing, whereas inference fluctuates with real-time user requests of varying length and complexity. Because GPU power supply units (PSUs) rely on high-frequency switch-mode converters \cite{gyangGridintegratedDataCenters2026}, EMT modeling is necessary to resolve sub-cycle load ramps, converter current harmonics, and terminal voltage/frequency sensitivity that phasor-domain RMS models smooth out.

\subsection{Uninterruptible Power Supply and Storage}

Coupled with a BESS, the UPS isolates critical IT loads from grid disturbances across three operating modes \cite{chenElectricityDemandGrid2026a}: double-conversion mode, where grid AC power is rectified and re-inverted into tightly regulated AC voltage; bypass mode, where IT loads connect directly to the grid during maintenance or high-efficiency operation; and battery backup mode, where the BESS supplies the IT loads during severe grid outages. EMT simulation is essential to capture transfer switch transitions, DC-link capacitor voltage transients, converter switching dynamics, and VRT/FRT control responses during faults.

\subsection{IT Load Power Electronics Configuration}
\label{sec: power elec}

AIDCs primarily employ three power electronics architectures: centralized UPS, distributed rack-level storage, and emerging DC distribution systems \cite{gyangGridintegratedDataCenters2026}. Centralized topologies are most common and isolate multi-rack IT loads, but their sub-cycle transfer logic and converter switching require EMT simulation to resolve high-speed mode transitions. In contrast, distributed designs eliminate double-conversion stages to reduce costs, directly exposing IT power supplies to transient voltage events. Emerging direct-current distribution eliminates output inverters to enhance power density \cite{gyangGridintegratedDataCenters2026}, introducing fast DC dynamics, resonance, and control interactions that RMS models cannot capture. Capturing these fast switching dynamics, transfer latencies, and disturbance vulnerabilities makes EMT modeling indispensable.

\subsection{Cooling Systems}
\label{sec: cooling}

Because rack power densities in AIDCs exceed 100 kW \cite{gyangGridintegratedDataCenters2026}, conventional fan-based air cooling is paired with chiller-based liquid cooling, such as direct-to-chip cold plates or liquid immersion, driven by variable frequency drives (VFDs) to limit motor inrush currents. From an EMT perspective, VFD front-end rectifiers inject non-sinusoidal harmonics, while IM fans and chillers exhibit distinct slip, stalling, and post-fault re-acceleration dynamics during voltage sags. Furthermore, coupling electrical cooling demand with rack thermal inertia is critical for evaluating sustained disturbance recovery.

\subsection{Grid Interface, Onsite Generation, and Site Controls}
\label{sec: generation}

Step-down transformers interface the AIDC with the utility grid at the point of interconnection (POI) and are safeguarded by protective relays. Facilities also integrate medium-voltage (MV) utility-scale gas turbines to offset grid demand alongside low-voltage (LV) backup diesel generators \cite{PNNLModel}. Finally, a supervisory site control system coordinates workload throttling, cooling curtailment, and generator dispatch. EMT modeling of these elements is required to capture transformer core saturation and inrush currents, protective relay trip coordination, generator synchronization transients, and control communication latencies during severe grid events. Together, these dynamics define the baseline architectural features examined across the open-source PSCAD models in Section \ref{sec: models} and the 11 evaluation categories established in Section \ref{sec: framework}.

\section{Overview of Open-Source EMT AIDC Models}
\label{sec: models}
\begin{table*}[t!]
    \caption{Open Source EMT AIDC Models in PSCAD}
    \centering
    \setlength{\tabcolsep}{5pt}
    \begin{tabularx}{\linewidth}{c X l l c}
        \toprule
        \textbf{Model \#} & \textbf{Brief Description} & \textbf{Institution} & \textbf{Last Updated} & \textbf{Test Cases} \\
        \midrule
        1 & Dynamic Modeling of AI Data Center Load in PSCAD \cite{ERCOTModel} & ERCOT/Texas A\&M & 05/2026 & 3 \\
        2 & Electromagnetic Transient Modeling of Data Centers \cite{NLRModel} & NLR & 08/2026 & 4 \\
        3 & Electromagnetic Transient Modeling of Large Data Centers for Grid-Level Studies \cite{PNNLModel} & PNNL & 08/2026 & 10 \\
        4 & Generic EMT Modeling of Data Center Load \cite{DartmouthModel} & Dartmouth College & 10/2026 & 2 \\
        \bottomrule
    \end{tabularx}
    
    \label{tab:emt models}
\end{table*}

Power System Computer Aided Design (PSCAD/EMTDC) is a graphical EMT simulation platform developed by Manitoba Hydro International that solves time-domain differential equations via its EMTDC engine \cite{pscad}. As of October 25, 2026, a survey of publicly available repositories identified four open-source PSCAD EMT AIDC models (Table \ref{tab:emt models}), referred to in our paper as Models 1–4 \cite{ERCOTModel}, \cite{NLRModel}, \cite{PNNLModel}, \cite{DartmouthModel}.

Model 1 from ERCOT and Texas A\&M University uses a modular, bottom-up architecture with averaged power electronic converter models to represent a double-conversion UPS facility, VFD-driven motor cooling, and aggregated AC/DC computing loads. Its three test cases demonstrate UPS operating mode transitions, voltage ride-through, and time-varying compute load fluctuations. Model 2 from the National Laboratory of the Rockies (NLR) represents a data center built from scaled 33 kW IT rack power shelves (six 5.5 kW power supply units), a multi-mode AC UPS (double-conversion, eco, and single-conversion), cooling loads, and a 260 MW onsite gas turbine generator across four grid-connected and off-grid use cases for converter-driven stability and power quality studies. Model 3 from the Pacific Northwest National Laboratory (PNNL) provides the Data Center Model Library, comprising 10 test cases across centralized UPS, distributed rack-level UPS, and mixed distribution topologies. Designed for bulk grid-level interconnection studies, it scales up to 1.54 GW total active power (1,050 MW IT load), matching the largest operational AIDCs, and separates cooling loads into aggregated 0.2 MW chiller and 0.6 MW fan units. Model 4 from Dartmouth College implements a modular centralized-UPS data center model offering average, switching, and reduced-order converter representations alongside motor-driven cooling and synthetic workload profiles. Its two test cases evaluate full-facility grid coupling and trade-offs across converter modeling fidelity levels.
To evaluate component parameters on a consistent architectural basis, we selected one centralized-UPS test case per model: \mbox{\textit{TestCase_ITLoadFluctuation_Ave}} (Model 1), \mbox{\textit{UseCase3_DataCenter}} (Model 2), \mbox{\textit{DML_DC2_Central_UPS}} (Model 3), and \mbox{\textit{DC_EMT_demo}} (Model 4). Table \ref{tab:power_ratings} summarizes the base and current-scaled power ratings of each test case, all of which exceed ERCOT's 75 MW LCL threshold. Because the models represent different facility ratings and test scenarios, we audit their component-level representations rather than comparing raw simulation waveforms directly.

\begin{table}[t!]
  \caption{Power Ratings of Selected Test Cases}
  \label{tab:power_ratings}
  \centering
  \renewcommand{\arraystretch}{0.9}
  \begin{tabularx}{\columnwidth}{lXcc}
    \toprule
    \textbf{Model \#} & \textbf{Load} & \textbf{Base Rating} & \textbf{Scaled Power} \\
    \midrule
    \multirow{4}{*}{1} 
      & IT        & 1.5~MW  & 150~MW \\
      & Cooling   & 0.05~MW & 5~MW \\
      & Auxiliary & N/A     & N/A \\
      & \multicolumn{2}{l}{\textbf{Total Active Power}} & \textbf{155~MW} \\
    \midrule
    \multirow{4}{*}{2} 
      & IT        & 1.65~MW & 132~MW \\
      & Cooling   & 0.5~MW  & 40~MW \\
      & Auxiliary & 0.1~MW  & 8~MW \\
      & \multicolumn{2}{l}{\textbf{Total Active Power}} & \textbf{180~MW} \\
    \midrule
    \multirow{4}{*}{3} 
      & IT        & 1~MW    & 1050~MW \\
      & Cooling   & 0.2 \& 0.6~MW & 200 \& 200~MW \\
      & Auxiliary & 100~MVA & 100~MVA ($0.9$ lag) \\
      & \multicolumn{2}{l}{\textbf{Total Active Power}} & \textbf{1540~MW} \\
    \midrule
    \multirow{4}{*}{4} 
      & IT        & 0.15~MW & 60~MW \\
      & Cooling   & 0.5~MW  & 20~MW \\
      & Auxiliary & 11.18~MVA & 11.18~MVA ($0.89$ lag) \\
      & \multicolumn{2}{l}{\textbf{Total Active Power}} & \textbf{90~MW} \\
    \bottomrule
  \end{tabularx}
\end{table}

\section{Proposed Evaluation Framework}
\label{sec: framework}
Table \ref{tab: checklist} presents our software-agnostic, 41-point evaluation framework across 11 categories for auditing EMT AIDC models. We apply this framework to evaluate the component-level completeness of the four open-source PSCAD models.

\begin{table}[t!]
    \centering
    \caption{EMT AIDC Model Evaluation Framework}
    \label{tab: checklist}
    \newcommand{\rotcat}[1]{\rotatebox[origin=c]{90}{\textbf{\shortstack{#1}}}}
    \setlength{\tabcolsep}{4pt} 
\begin{tabularx}{\columnwidth}{c l X}
\toprule
\textbf{Cat.} & \textbf{ID} & \textbf{Specification} \\
\midrule
\multirow{4}{*}{\rotcat{Utility \&\\MV Bus}} 
 & A.1 & Dynamic voltage-/frequency-sensitive POI with impedance \\
 & A.2 & MV protection relays \& utility coordination schemes \\
 & A.3 & Simulation of voltage sags/swells, frequency excursions, \& AI-load-driven POI oscillations \\
\midrule

\multirow{5}{*}{\rotcat{Trans-\\formers}} 
 & B.1 & Inrush current \& core saturation phenomena \\
 & B.2 & On-load tap changer (OLTC) control behavior \\
 & B.3 & Protective relaying schemes for overcurrent, differential protection, \& under/overvoltage trips \\
 & B.4 & Interaction with fast-varying DC loads \& grid disturbances \\
\midrule

\multirow{2}{*}{\rotcat{IT}} 
 & C.1 & Time-varying load transitions \& rapid ramp capabilities \\
 & C.2 & Load sensitivity to voltage \& frequency variations \\
\midrule

\multirow{3}{*}{\rotcat{UPS}} 
 & D.1 & Dual-mode operation (grid-following \& grid-forming) \\
 & D.2 & Transition logic for choosing utility power \& backup \\
 & D.3 & Compliance with VRT requirements \\
\midrule

\multirow{4}{*}{\rotcat{Cooling\\Loads}} 
 & E.1 & Voltage \& frequency sensitivity of motor-driven components \\
 & E.2 & Staged startup, load shedding, \& curtailment control logic \\
 & E.3 & Thermal inertia effects influencing disturbance recovery times \\
 & E.4 & Coordination with facility management systems \\
\midrule

\multirow{3}{*}{\rotcat{BESS}} 
 & F.1 & Fast transient response \& dynamic control during grid events \\
 & F.2 & Compliance with IEEE Std 1547 \cite{IEEEStandardInterconnection2018} \\
 & F.3 & State-of-charge management \& site controller interactions \\
\midrule

\multirow{7}{*}{\rotcat{Generator\\Models}} 
 & G.1 & Startup delays \& ramp rates \\
 & G.2 & Load acceptance dynamics \& backup power connection times \\
 & G.3 & V, I, \& f protective relays \\
 & G.4 & Synchronization \& reconnection mechanisms \\
 & G.5 & Automatic generation control \\
 & G.6 & Automatic voltage regulation \\
 & G.7 & Isochronous control, droop control, \& load-sharing schemes \\
\midrule

\multirow{5}{*}{\rotcat{Protection\\Systems}} 
 & H.1 & Protective relay logic with thresholds for V, I, \& f \\
 & H.2 & Single- \& three-pole circuit breaker models with trip time delays \\
 & H.3 & Relay coordination \& cascading trip sequences \\
 & H.4 & Instrument transformer saturation effects \& harmonic filtering \\
\midrule

\multirow{5}{*}{\rotcat{Higher-Level\\Control}} 
 & I.1 & Custom control sequences for breaker operation, load/DER dispatch, \& demand response \\
 & I.2 & Real-time communication \& monitoring interfaces \\
 & I.3 & Fault detection, system isolation, \& black-start control strategies \vspace{2pt} \\
\midrule

\multirow{5}{*}{\rotcat{Transfer\\Switches}} 
 & J.1 & Fast transfers between independent utility \& backup sources \\
 & J.2 & Sensing- \& logic-driven transfer initiation \\
 & J.3 & Realistic transfer time, transient response, \& interruption characteristics \\
 & J.4 & Voltage synchronization, inrush control, \& load acceptance \\

\midrule
\multirow{4}{*}{\rotcat{Ours}}
 & K.1 & Compliance with FRT requirements \\
 & K.2 & Separate modeling of fan- \& chiller-based cooling loads \\
 & K.3 & Temperature-dependent cooling power response \\
 & K.4 & DC-side IT load representation \\ 
\bottomrule

\end{tabularx}
\end{table}
Categories A-J (37 points) are synthesized from \cite{talukdarModelingFrameworkData2026}, which define required component representations across ten modeling layers alongside recommended PSCAD implementations. To address emerging trends identified in Sections \ref{sec: intro} and \ref{section:Data Center Design}, we introduce four additional requirements in Category K:
(K.1) FRT Compliance---Incorporating frequency ride-through (FRT) modules into UPS controls to reflect evolving LCL interconnection mandates; (K.2) Disaggregated Cooling Loads---Separating fan- and chiller-based cooling subsystems to capture their distinct electrical and operational characteristics; (K.3) Temperature-Dependent Cooling---Coupling cooling power consumption to ambient or rack operating temperatures; and (K.4) DC IT Load Representation---Modeling IT equipment as DC loads rather than aggregated AC equivalents that abstract power electronic converter stages.

\section{Model Evaluation Results}

Table \ref{tab:section_scores} summarizes the evaluation of the four open-source EMT models against our 41-point framework, with total scores ranging from 13 to 21. Subsections \ref{sec: A} through \ref{sec: K} correspond to Categories A-K in Table \ref{tab: checklist}, while Subsection \ref{sec: L} synthesizes facility-wide disturbance ride-through capabilities. Across the suite, the models provide strong EMT representations of IT load profiles, UPS, and transfer switches. Because these releases were scoped for bulk grid-level stability studies, thermodynamic cooling feedback, internal protection relaying, and site-level communications remain outside their current scope, highlighting clear priorities for future model extensions. Models can also benefit from detailed disturbance ride-through control implementation for disturbance studies.

\subsection{Utility Interface and Medium-Voltage Bus}
\label{sec: A}

All four models establish POIs with configurable grid impedances and short-circuit ratios. Models 1-3 include standard North American HV bus voltages, whereas Model 4 starts at the MV level; all models include standard MV and 480 V LV buses. Model 3 represents the grid via a scaled synchronous generator, step-up transformer, and series impedances, while Models 1, 2, and 4 use non-ideal three-phase voltage sources. Adding MV protective relays, utility coordination schemes, and frequency excursion test cases would enable broader compliance studies.

\subsection{Transformers}
All four models use PSCAD three-phase, two-winding transformer blocks equipped with inrush and core saturation representations (B.1) that capture EMT interactions (B.4). Although saturation is toggled off in the default test cases, likely due to limited nameplate data, it can be readily enabled once parameters are specified.

\subsection{Information Technology Loads}

Rapid IT load transitions are modeled via rack-level variable resistances driven by empirical cluster telemetry in Model 1 (MIT Supercloud Dataset v1 \cite{samsiMITSupercloudDataset2021}), synthetic AI profiles in Models 2 and 3 (ON.energy and EdgeTunePower \cite{edgetunepowerETPAILoad}), and square, sinusoidal, and ramp waveforms in Model 4. Models 2 and 4 retain power electronic converter interfaces that capture voltage and frequency sensitivity, whereas Models 1 and 3 abstract the converter stage. All models use current scalers to scale single-rack loads to multi-megawatt ratings; future studies can verify aggregation linearity during fast transients.

\subsection {Uninterruptible Power Supply}
All four models provide detailed dual-mode UPS representations with automated utility-to-backup transition logic. Models 1, 3, and 4 support programmable VRT curves, and Model 1 additionally incorporates configurable FRT settings.
\begin{table}[t!]
  \caption{Evaluation Scores Across Models}
  \label{tab:section_scores}
  \centering
  \setlength{\tabcolsep}{2.8pt}
  \begin{tabularx}{\columnwidth}{X *{11}{c} c}
    \toprule
    \textbf{Category} & \textbf{A} & \textbf{B} & \textbf{C} & \textbf{D} & \textbf{E} & \textbf{F} & \textbf{G} & \textbf{H} & \textbf{I} & \textbf{J} & \textbf{K} & \textbf{Sum} \\
    \textit{Max Score} & \textit{(/3)} & \textit{(/4)} & \textit{(/2)} & \textit{(/3)} & \textit{(/4)} & \textit{(/3)} & \textit{(/7)} & \textit{(/4)} & \textit{(/3)} & \textit{(/4)} & \textit{(/4)} & \textit{(/41)} \\
    \midrule
    \textbf{Model 1} & 1 & 2 & 1 & 3 & 1 & 2 & 0 & 0 & 0 & 2 & 2 & \textbf{14} \\
    \textbf{Model 2} & 1 & 2 & 2 & 2 & 1 & 2 & 7 & 0 & 0 & 2 & 1 & \textbf{20} \\
    \textbf{Model 3} & 1 & 2 & 1 & 3 & 1 & 1 & 7 & 1 & 0 & 3 & 1 & \textbf{21} \\
    \textbf{Model 4} & 1 & 2 & 2 & 3 & 1 & 1 & 0 & 0 & 0 & 2 & 1 & \textbf{13} \\
    \bottomrule
  \end{tabularx}
\end{table}
\subsection{Cooling Loads}
Cooling systems are represented using voltage- and frequency-sensitive IMs. Models 1 and 3 used electrical parameters identical to PSCAD IM library defaults without rescaling for base voltage and power, while Models 2 and 4 aggregate cooling into 0.5 MW IM equivalents. Because cooling demand is currently decoupled from IT utilization, integrating VFD curtailment, thermal inertia, and facility coordination will improve post-disturbance recovery fidelity.

\subsection{Battery Energy Storage System}

All models capture fast BESS transient response and dynamic power control. Models 1 and 2 track SOC dynamically, whereas Models 3 and 4 hold SOC constant, a practical simplification over short EMT windows. Additional communication modeling will allow models to meet IEEE Std 1547 \cite{IEEEStandardInterconnection2018}.

\subsection{Generator Models}

Models 2 and 3 incorporate onsite generation—a 260 MW gas turbine (300 MVA, 13.2 kV synchronous machine) in Model 2 and a reciprocating engine with an autosynchronizer in Model 3—whereas Models 1 and 4 represent grid-supplied facilities. As noted by Model 3's developers \cite{PNNLModel}, more sophisticated generation models are needed for plant-level studies at the design stage. This could include adding governor droop, automatic generation control, and load-sharing logic.

\subsection{Protection Systems}
Current protection modeling focuses primarily on the UPS interface. Model 3 extends coverage to the wider facility using configurable single- and three-phase breakers with undervoltage and voltage-unbalance relays. Building on Model 3 by adding overcurrent and frequency relays, time-graded coordination, and instrument transformer saturation will support internal fault and cascading trip analysis.

\subsection{Higher-Level Control Systems}
Site-level supervisory controls remain an area for development across all four models. Extending existing utility-fault test cases with coordinated load dispatch, demand response, communication latencies, and internal fault isolation will enable full facility-level disturbance studies.

\subsection{Automatic/Static Transfer Switches}
All models link transfer switches to UPS sensing and transition logic. Model 3 further incorporates voltage synchronization, inrush control, and staged load acceptance for onsite generation. Adding realistic mechanical/static transfer delays and arc-interruption dynamics across all models would refine sub-cycle transfer accuracy.

\subsection{Additional Evaluation Points}
\label{sec: K}

Across the four proposed criteria in Category K, Model 1 incorporates UPS FRT settings (K.1), Model 3 disaggregates fan and chiller cooling loads with separate scaling factors (K.2), and Models 1, 2, and 4 include DC-side IT load representations (K.4). None of the models currently couple temperature dynamics to cooling power consumption (K.3), leaving electro-thermal feedback as an important next step.

\subsection{Disturbance Ride-Through Control}
\label{sec: L}

While Models 1, 3, and 4 capture local UPS VRT behavior (D.3) and Model 1 includes FRT compliance (K.1), the absence of site-level supervisory control (Category I) prevents end-to-end simulation of coordinated IT throttling, cooling curtailment, generation dispatch, and UPS transfers. Integrating facility-wide ride-through controls is therefore essential for studying and mitigating large-scale load transfer events such as those described in Section \ref{sec: intro}.

\section{Conclusion}

This paper introduced a software-agnostic, 41-point fidelity framework spanning 11 categories to audit the component-level completeness of EMT AIDC models, integrating emerging requirements for frequency ride-through, disaggregated and temperature-dependent cooling, and DC IT load representations. Applying this framework to the four publicly available PSCAD/EMTDC AIDC models shows that current releases provide strong foundational representations of IT load fluctuations, UPS dynamics, and transfer switches for bulk grid-level stability studies. Extending these open-source libraries with electro-thermal cooling feedback, internal protective relaying, and supervisory site-level controls, alongside standardized cross-model output and runtime benchmarking, will enable comprehensive protection coordination and facility-wide ride-through compliance studies.

\ifCLASSOPTIONcaptionsoff
  \newpage
\fi

\printbibliography

@article{chenElectricityDemandGrid2026a,
  title = {Electricity Demand and Grid Impacts of {{AI}} Data Centers: {{Challenges}} and Prospects},
  shorttitle = {Electricity Demand and Grid Impacts of {{AI}} Data Centers},
  author = {Chen, Xin and Wang, Xiaoyang and Colacelli, Ana and Lee, Matt and Xie, Le},
  date = {2026-09-15},
  journaltitle = {Nexus},
  shortjournal = {Nexus},
  volume = {3},
  number = {3},
  pages = {100162},
  issn = {2950-1601},
  doi = {10.1016/j.ynexs.2026.100162},
  url = {https://www.sciencedirect.com/science/article/pii/S2950160126000458},
  urldate = {2026-11-04},
}

@software{DartmouthModel,
  title = {Generic {{EMT Modeling}} of {{Data Center Load}} v0 ({{DC}}\_{{EMTv0}})},
  author = {Zhao, Xingyu and Tang, Yingyi and Zhao, Junbo},
  date = {2026},
  url = {https://github.com/xingyuz-phd/Generic-EMT-Modeling-of-Data-Center-Load},
  organization = {Dartmouth College}
}

@dataset{edgetunepowerETPAILoad,
  title = {{{ETP AI Load Profile Generator}}},
  author = {{EdgeTunePower}},
  url = {https://www.edgetunepower.com/copy-of-data-center-and-ibr-assessment}
}

@misc{epochaiAIDataCenters2026,
  title = {{{AI}} Data Centers},
  author = {{Epoch AI}},
  date = {2026-11},
  url = {https://epoch.ai/data/ai-data-centers}
}

@report{ERCOTModel,
  title = {Dynamic {{Modeling}} of {{AI Data Center Load}} in {{PSCAD}}},
  author = {Wang, Xiaoyang and Chen, Xin and Enjeti, Prasad and Lin, Fang Chen and Yazdanpanah, Ali and Rose, Jonathan and Cheng, Yunzhi and Quedan, Amro and Kansal, Prashant},
  date = {2026-04},
  pages = {105},
  institution = {TAMU-ERCOT},
  url = {https://www.ercot.com/files/docs/2026/04/24/Dynamic-Modeling-of-AI-Data-Center-Load-in-PSCAD-Manual.pdf}
}

@online{ercotNOGRR282LargeComputational2026,
  title = {{{NOGRR282}} - {{Large Computational Load Ride-Through Requirements}}},
  author = {{ERCOT}},
  date = {2026-09-07},
  location = {Texas},
  url = {https://www.ercot.com/mktrules/issues/NOGRR282#summary},
  urldate = {2026-11-05},
  organization = {Board Priority - Large Computational Load Ride-Through Requirements},
}

@article{favuzzaComparativeModelingAnalysis2024,
  title = {Comparative Modeling and Analysis of EMT and Phasor RMS Grid-Forming Converters under Different Power System Dynamics},
  author = {Favuzza, Salvatore and Musca, Rossano and Zizzo, Gaetano and Sa'ed, Jaser A.},
  date = {2024},
  journaltitle = {IEEE Transactions on Industry Applications},
  volume = {60},
  number = {2},
  pages = {3613--3624},
  doi = {10.1109/TIA.2023.3339108}
}

@article{gyangGridintegratedDataCenters2026,
  title = {Grid-Integrated Data Centers: {{Architectures}}, Modeling, Interconnection Requirements, Challenges, and Opportunities},
  author = {Gyang, Pam Paul and Chakraborty, Pratyush and Meegahapola, Lasantha and Yu, Xinghuo},
  date = {2026},
  journaltitle = {IEEE Open Journal of the Industrial Electronics Society},
  volume = {7},
  pages = {1080--1112},
  doi = {10.1109/OJIES.2026.3706771}
}

@article{IEEEStandardInterconnection2018,
  title = {{{IEEE Standard}} for {{Interconnection}} and {{Interoperability}} of {{Distributed Energy Resources}} with {{Associated Electric Power Systems Interfaces}}},
  shorttitle = {{{IEEE Std}} 1547-2018},
  date = {2018},
  journaltitle = {IEEE Std 1547-2018 (Revision of IEEE Std 1547-2003)},
  pages = {1--138},
  publisher = {IEEE},
  location = {New York, NY, USA},
  doi = {10.1109/IEEESTD.2018.8332112},
  langid = {american},
}

@report{ieso2026largecomputational,
  type = {Technical Requirements},
  title = {Technical Requirements for Large Computational Loads Connecting to the Ontario Power System},
  author = {{IESO}},
  date = {2026-05-01},
  institution = {Independent Electricity System Operator (IESO)},
  location = {Toronto, ON, Canada},
  url = {https://ieso.ca/-/media/Files/IESO/Document-Library/engage/large-computational-loads/LargeComputationalLoads-20260501-Technical-Requirements.pdf},
  eid = {Version 1.0}
}

@article{koWideareaPowerSystem2026,
  title = {Wide-Area Power System Oscillations from Large-Scale {{AI}} Workloads},
  author = {Ko, Min-Seung and Zhu, Hao},
  date = {2026},
  journaltitle = {IEEE Transactions on Power Systems},
  volume = {41},
  number = {5},
  pages = {3908--3921},
  doi = {10.1109/TPWRS.2026.3685506}
}

@report{NLRModel,
  type = {Technical Report},
  title = {Electromagnetic {{Transient Modeling}} of {{Data Centers}}},
  author = {Shah, Shahil and Yan, Weihang},
  date = {2026-08},
  pages = {49},
  institution = {National Laboratory of the Rockies},
  location = {Golden, CO.},
  url = {https://www.nlr.gov/docs/fy26osti/101189.pdf},
  urldate = {2026-11-04},
  eid = {NLR/TP-5D00-101189},
  langid = {american}
}

@report{PNNLModel,
  title = {Electromagnetic {{Transient Modeling}} of {{Large Data Centers}} for {{Grid-Level Studies}}: {{Beta Release}}},
  shorttitle = {Electromagnetic {{Transient Modeling}} of {{Large Data Centers}} for {{Grid-Level Studies}}},
  author = {Ross, Brett A. and Nekkalapu, Sameer and Lyu, Xue and Follum, James D.},
  date = {2026-08-14},
  institution = {Pacific Northwest National Laboratory (PNNL), Richland, WA (United States)},
  doi = {10.2172/3422378},
  url = {https://www.osti.gov/biblio/3422378},
  urldate = {2026-11-04},
  eid = {PNNL-38817 Beta},
  langid = {english}
}

@manual{pscad,
  type = {manual},
  title = {{{PSCAD}}/{{EMTDC}}: {{Electromagnetic}} Transients Program Including {{DC}} Systems},
  author = {{Manitoba Hydro International Ltd.}},
  publisher = {Manitoba Hydro International Ltd.},
  location = {Winnipeg, MB, Canada},
  url = {https://www.pscad.com}
}

@inproceedings{samsiMITSupercloudDataset2021,
  title = {The {{MIT}} Supercloud Dataset},
  booktitle = {2021 {{IEEE}} High Performance Extreme Computing Conference ({{HPEC}})},
  author = {Samsi, Siddharth and Weiss, Matthew L and Bestor, David and Li, Baolin and Jones, Michael and Reuther, Albert and Edelman, Daniel and Arcand, William and Byun, Chansup and Holodnack, John and Hubbell, Matthew and Kepner, Jeremy and Klein, Anna and McDonald, Joseph and Michaleas, Adam and Michaleas, Peter and Milechin, Lauren and Mullen, Julia and Yee, Charles and Price, Benjamin and Prout, Andrew and Rosa, Antonio and Vanterpool, Allan and McEvoy, Lindsey and Cheng, Anson and Tiwari, Devesh and Gadepally, Vijay},
  date = {2021},
  pages = {1--8},
  doi = {10.1109/HPEC49654.2021.9622850}
}

@article{shahriarpramanikAssessingDynamicBehaviors2025,
  title = {Assessing {{Dynamic Behaviors}} in {{Converter}}- {{Dominated Power Systems}} via {{RMS}} and {{EMT Simulations}}: {{A Study}} of {{Hawaii}}’s {{NELHA Microgrid}}},
  shorttitle = {Assessing {{Dynamic Behaviors}} in {{Converter}}- {{Dominated Power Systems}} via {{RMS}} and {{EMT Simulations}}},
  author = {Shahriar Pramanik, Abrar and Sepasi, Saeed and Griswold, Brian and Adegboyega, Adewale},
  date = {2025},
  journaltitle = {IEEE Access},
  volume = {13},
  pages = {83881--83896},
  issn = {2169-3536},
  doi = {10.1109/ACCESS.2025.3568782},
  url = {https://ieeexplore.ieee.org/document/11000107/},
  urldate = {2026-11-04},
}

@report{talukdarModelingFrameworkData2026,
  title = {Modeling {{Framework}} for {{Data Center}}},
  author = {Talukdar, Saurav and Marti, Shilpa and Prabakar, Kumaraguru and Vaidhynathan, Deepthi},
  date = {2026-02-13},
  institution = {National Laboratory of the Rockies (NLR), Golden, CO (United States)},
  doi = {10.2172/3019724},
  url = {https://www.osti.gov/biblio/3019724},
  urldate = {2026-11-05},
  eid = {NLR/TP-2C00-97716},
  langid = {english},
}

@misc{wharton2026dominion,
  title = {July 22, 2026 Dominion Load Transfer Event},
  author = {Wharton, Matthew},
  date = {2026-07-31},
  url = {https://www.pjm.com/-/media/DotCom/committees-groups/subcommittees/sos/2026/20260730/20260730-item-06---july-22-dominion-load-transfer-event---presentation.pdf},
  howpublished = {Presentation to the PJM System Operations Committee},
  organization = {PJM Interconnection}
}

\end{document}